\documentclass{article}

\usepackage{arxiv}

\usepackage[utf8]{inputenc} 
\usepackage[T1]{fontenc}    
\usepackage{hyperref}       
\usepackage{fancyref}
\usepackage{url}            
\usepackage{booktabs}       
\usepackage{amsfonts}       
\usepackage{nicefrac}       
\usepackage{microtype}      
\usepackage{lipsum}
\usepackage{graphicx}
\usepackage[numbers]{natbib}
\graphicspath{ {./images/} }
\title{What Do End-Users Really Want? Investigation of Human-Centered XAI for Mobile Health Apps}

\author{
 Katharina Weitz \\
 Chair for Human-Centered AI\\
 University of Augsburg\\
 Universit\"atsstraße 6a  \\
  86159 Augsburg \\
  \texttt{katharina.weitz@uni-a.de} \\
   \And
 Alexander Zellner \\
 Chair for Human-Centered AI\\
  University of Augsburg\\
  Universit\"atsstraße 6a  \\
  86159 Augsburg \\
  \texttt{alexander.zellner@informatik.uni-augsburg.de} \\
\And
  Elisabeth Andr{\'e}\\
 Chair for Human-Centered AI\\
  University of Augsburg\\
  Universit\"atsstraße 6a  \\
  86159 Augsburg \\
  \texttt{elisabeth.andre@uni-a.de} \\
}

\begin{document}
\maketitle
\begin{abstract}
In healthcare, AI systems support clinicians and patients in diagnosis, treatment, and monitoring, but many systems' poor explainability remains challenging for practical application.
Overcoming this barrier is the goal of explainable AI (XAI). However, an explanation can be perceived differently and, thus, not solve the black-box problem for everyone. The domain of Human-Centered AI deals with this problem by adapting AI to users. 
We present a user-centered persona concept to evaluate XAI and use it to investigate end-users preferences for various explanation styles and contents in a mobile health stress monitoring application. The results of our online survey show that users' demographics and personality, as well as the type of explanation, impact explanation preferences, indicating that these are essential features for XAI design. We subsumed the results in three prototypical user personas: power-, casual-, and privacy-oriented users. Our insights bring an interactive, human-centered XAI closer to practical application.

\end{abstract}

\maketitle

\section{Introduction}
\begin{figure}
  \includegraphics[width=\textwidth]{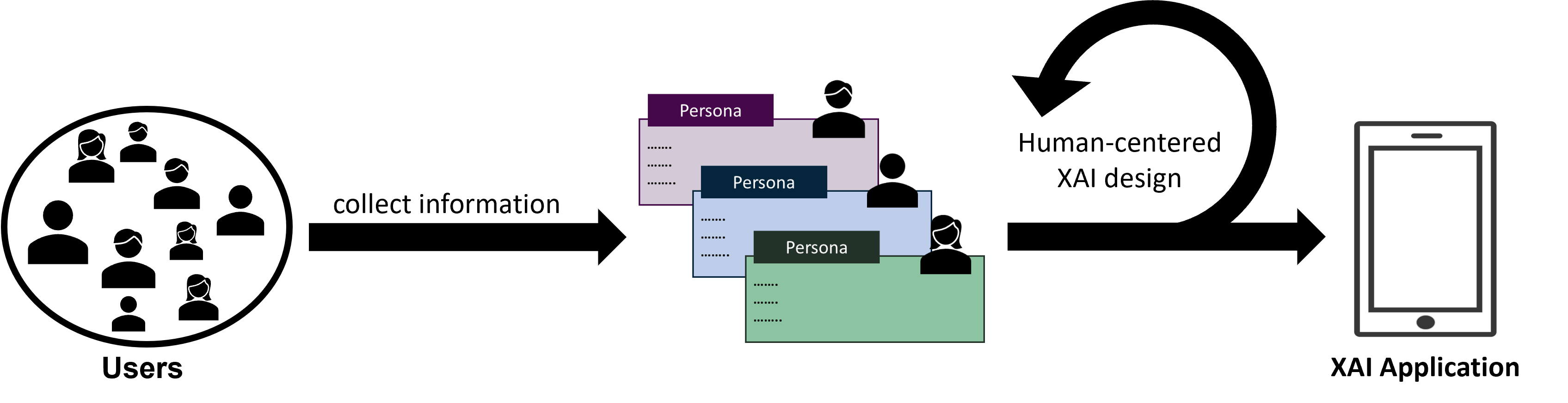}
  \caption{Including stakeholder into the design of human-centered XAI by creating personas based on data of real users}
  \label{fig:teaser}
\end{figure}
AI has become ubiquitous, and we have become used to AI as a decision support system. However, such decisions are not always equally impactful for our lives. An AI making movie or music recommendations
versus an AI diagnosing or not diagnosing life-threatening diseases have different impacts on our lives.
Knowing the reasons behind such important decisions is essential for not blindly trusting the AI.
The General Data Protection Regulation (GDPR) passed by the European Parliament in 2018
further supports this need for explanations \cite{EUdataregulations2018} and states in Art. 12 that the data of a subject has to be processed ``in a concise, transparent, intelligible and easily accessible form, using clear and plain language'' \cite{EUdataregulations2018}.
The area of Explainable Artificial Intelligence (XAI) aims to make AI
more understandable and transparent. XAI for ML focuses on supporting users to ``appropriate trust, and effectively manage'' AI \cite[p. 44]{gunningaha2019darpa}.
Despite this large body of research, a gap exists between what society needs and what researchers provide. \citet{miller2017beware} claims that developers mostly design explanations for other developers. A reason for this gap is that different people might require different kinds
of explanations \cite{arya2019one}. Following \citet{arya2019one} with their postulation, ‘one explanation does not fit all’ to provide satisfactory explanations to required fields, like the health sector, we need explanations to adapt to their receiver. A doctor will require a different explanation than a patient or an ML developer.
It requires the investigation of XAI to satisfy end-users, of which personalized explanations
will be one foothold. \citet{shneiderman2020trustworthyHCAI} suggests interactive explanations that enable greater user involvement and help users understand an ML system's behavior.
Therefore, besides a more algorithmic-driven evaluation of XAI, user-focused studies exploring user goals, knowledge, preferences, and intentions have gained increasing importance in XAI research. A personalized XAI system that adapts to the explanation recipient and fits their desires to explain the AI prediction appropriately can increase the practicality of XAI and allow its
usage beyond AI researchers towards becoming an integral part of everyday life. 

To come closer to this goal, we present a user-centered concept to evaluate XAI for practical applications using personas. This concept gains knowledge of end-users preferences regarding XAI and provides an approach to creating empirical-based personas representing end-user groups. Our approach supports XAI designers in gaining insights into their users' preferences and using this information to foster human-centered design (see \Fref{fig:teaser}). 
Based on our user-centered XAI persona concept, we are investigating in this paper end-users preferences for different explanation styles and contents in an online survey. From the collected data, we derive personas that describe prototypical end-users for mobile-health applications. 

\subsection{Related Work}
\subsection{Taxonomy of Explanations}
A variety of taxonomies can be used to classify XAI methods. One of the commonly used divides XAI techniques into model-agnostic and model-specific approaches.
Model-agnostic refers to algorithms like LIME \cite{ribeiro2016should} and SHAP \cite{lundberg2017unified} that are applied independently of the model’s characteristics so that they can be used for different AI models \cite{pedreschi2019meaningful}. 
In contrast, model-specific approaches are developed for specific ML methods, for example, the LRP approach \cite{bach2015pixel} and the Grad-CAM \cite{selvaraju2017grad} that work incredibly well on deep neural networks.
A diverse repertoire of XAI approaches has been developed
utilizing varying techniques. These broadly separate into four categories \cite{mothilal2020explaining, adadi2018peeking}:
\begin{itemize}
    \item \textbf{Visualisation} A natural way of bridging ML models’ complexity and algebraic nature is by using visualization techniques. For example, \citet{cortez2011opening} presents a portfolio of visualizations to improve the explainability of black-box building upon the Global SA technique.
    \item \textbf{Feature-based} This technique estimates the importance, influence, or relevance of individual features on the prediction. For example, in \citet{lundberg2017unified}, the authors present the SHAP (SHapley Additive exPlanations) framework. It calculates an additive feature importance score taking additional properties such as accuracy and consistency discerning from their antecedents.
    \item \textbf{Knowledge extraction} Learning algorithms modify cells in the hidden layer of a model. The task of knowledge-based explanation is to extract, in a comprehensible representation, the knowledge acquired by the network during training phases \cite{adadi2018peeking}. A commonly seen approach is a rule-based explainer, as the rule extractions proposed by \citet{hailesilassie2016rule} or aLIME \cite{ribeiro2016should} providing if-then rules in a model-agnostic manner.
    \item \textbf{Example-based} This represents a unique technique among the previously mentioned. Example-based explanations are model-agnostic since they improve the interpretability of any ML model. However, they interpret a model based on the present dataset, not on features
    or model transformations. One of the most promising approaches is the explanation through counterfactuals \cite{mothilal2020explaining,wachter2017counterfactual}.
\end{itemize}

The presented classification of techniques is based on active research. Therefore, new methods are proposed, and existing ones are extended. 

Besides the development of XAI approaches, researchers like \citet{doshi2017towards} point to the necessity of evaluating XAI to investigate the usefulness of these methods. They proposed a taxonomy for evaluating XAI. In their three-step approach, they start with \textit{functionally-grounded evaluation} with proxy tasks without humans. In these proxy tasks, a formal explanation quality is evaluated. The following two steps, namely \textit{human-grounded evaluation} and \textit{application-grounded evaluation}, involve users and focus either on simple or real-world application tasks. \citet{doshi2017towards} highlight that from step to step, the evaluation approach gets more cost-intensive and more specific due to the application of XAI. 

We focus on a human-grounded evaluation as proposed by \citet{doshi2017towards}. In doing so, we investigate end-users preferences for a mobile health stress monitoring app. In accordance with \citet{doshi2017towards}, we investigate the impact of different types and contents of an explanation in a mock-up setup. This benefits us to focus on the explanation design without being limited by the outcome or performance of current ML models. 

\subsection{XAI for Medical Decision Support}
\begin{figure}[ht]
  \centering
  \includegraphics[width=\linewidth]{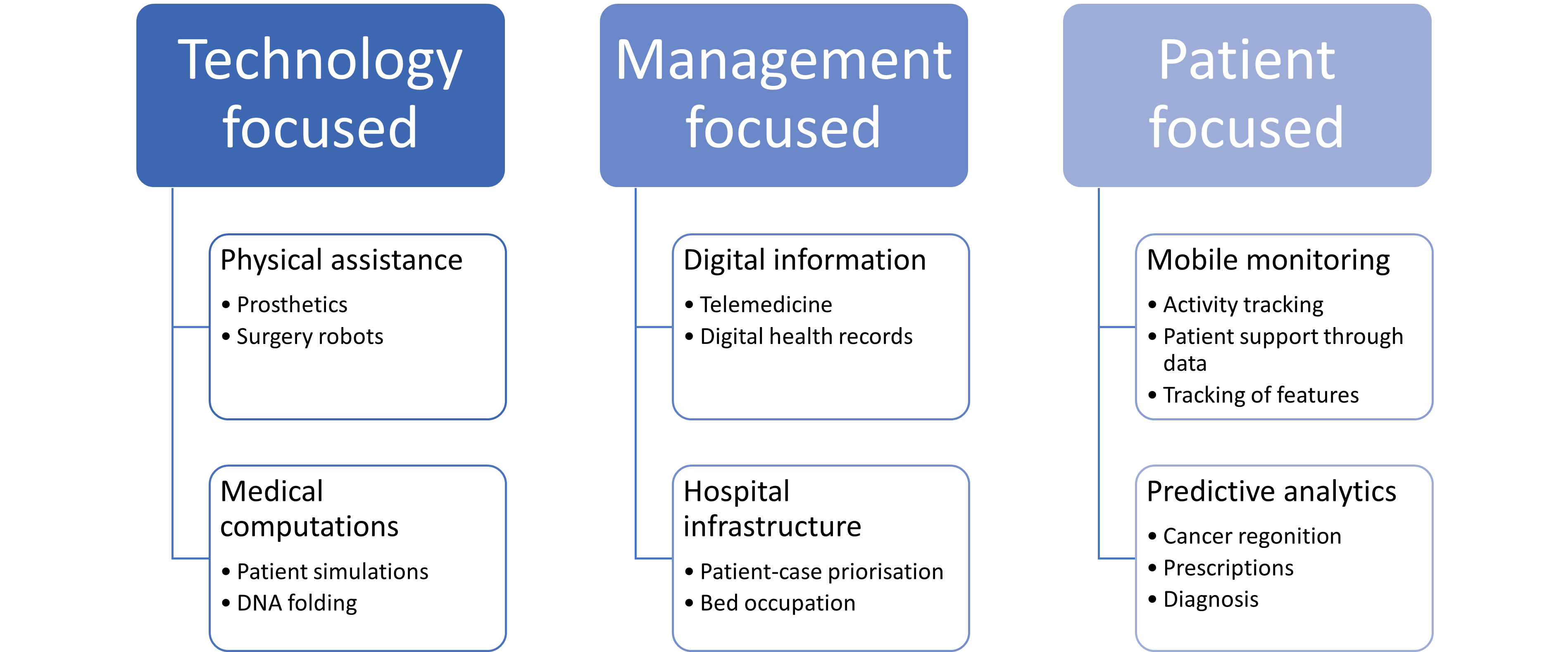}
  \caption{Overview about different purposes of AI in medical applications}
  \label{fig:OverviewMedicalApplicationOfAI}
\end{figure}

In the early days of AI, rule-based systems like MYCIN \cite{shortliffe1975computer} provided a first decision support system for clinicians. Since then, ML has gained momentum in
the medical field is leading to significant advancement in automated diagnosis and forecasting \cite{esteva2017dermatologist,nam2019development}.
The platform ‘grand-challenge.org’\footnote{\url{https://grand-challenge.org/}} lists over 328 different classification and image analysis tasks with medical backgrounds and the corresponding AI solutions. 
However, critical voices highlight that AI helped in experimental settings but failed in practice due to the poor fit to real-world requirements \cite{yang2019unremarkable}.
One of them is the black-box problem of ML, which increased the research community's interest in XAI and interpretability within the medical field. The question of responsibility and risk management are closely coupled to the usage of AI in medical contexts. Since when diagnoses are made, lives are at stake, leaving such a decision to an opaque ML model would be irresponsible on many levels.
Therefore, many have been dedicated to exploring and improving the transparency and explainability of AI in medicine \cite{tjoa2020medicalXAI}. In most of the ongoing research, diagnosis and classification tasks are the most apparent use cases when applying AI in the medical domain.
In contrast, a system developed by \citet{kostopoulos2017stress} detects the user’s stress level by analyzing smartphone data. This trend of registering and analyzing body conditions gained increasing attention with smart wearables allowing more accessible and continuous tracking. Consequently, users can now track
their heart rate, stress level, or physical activity throughout the day. According to recent studies, smart-wearables will significantly impact digital health, \citet{perez2021wearables} categorizes them in a subclass called \textit{mobile health}, standing out through their monitoring capabilities.
In \Fref{fig:OverviewMedicalApplicationOfAI} we summarize common digital health use-cases found in literature and structure them further according to similarities in their goal, applicability, or depth of intervention in medicine.

The concrete health use cases are thus summarized in six different categories. Which are then, in turn, further condensed into three dimensions:
\begin{itemize}
    \item \textbf{Technology-focused}: Here, the algorithm and the intended use are dominant. These are nearly exclusively used by domain experts (e.g., clinicians), and it is vital to understand their explainability, to comprehend ongoing processes on a technical level.
    \item \textbf{Management focused}: This targets the whole organization and administering digital health
    sectors instead of individual patients. The user group includes health professionals as well as
    administrative staff. Thus, explainability helps to retrace decisions, e.g., when it has to be explained to outside parties.
    \item \textbf{Patient-focused}: Revolves around the patient, either in providing diagnosis and assisting doctors in the process or monitoring the patient’s body, not necessarily involving a physician. This firm patient reference offers good preconditions for the envisaged personalisation approaches. 
\end{itemize}

Our paper focuses on the third dimension, \textit{patient focused} AI-based health applications. Here we investigate user preferences in a \textit{mobile monitoring} task for stress recognition.

\subsection{Human-Centered XAI}
Authors like \citet{miller2019explanation} and \citet{hoffman2018MetricsXAI} stressed that the same explanation could affect recipients differently due to cognitive and social variations and desired context. For example, an explainee's goal could be to understand why this circumstantial explanation was received or how the AI came to this conclusion. Hence, one explanation can not satisfy these heterogeneous goals. The omission of addressing these needs and interests can decrease or prevent the success of XAI applications in practice. \citet{arrieta2020XAIconcepts} designate the approximation dilemma to this issue; explanations must match the audience's requirements. This is addressed by Human-Centered AI (HCAI). HCAI provides a perspective on AI that highlights the necessity to take stakeholders' abilities, beliefs, and perceptions into the design of AI applications \cite{riedl2019human}.

A promising attempt to reach human-centered XAI is the personalisation of explanations described by \citet{schneider2019personalized}. Personalisation is incorporated by adapting, among other things, to the explainees’ knowledge, intent, and preferences. First attempts are shown in \citet{schneider2019personalized} and \citet{arya2019one} focus on static explanations often supported by text. So far, interactive visual approaches – potentially better suited for personalisation – have not been the focus of research or are limited to concepts and techniques closely related to AI developers \cite{miller2017beware}. \citet{shneiderman2020trustworthyHCAI} highlight that interactive explanations could enable greater user involvement and thus help users better understand an ML system's behaviour than static explanations.

Finally, it is worth noting that most of the XAI techniques are static in their explainability \cite{arya2019one}. They do not
change in response to feedback or reactions from the receiver. In contrast, an interactive explanation
allows consumers to immerse themselves in the explanation, e.g., ask questions. \citet{arya2019one, shneiderman2020trustworthyHCAI} highlight the importance of interactive explanations, especially for end-users, but missing evaluations.
The existing static explanations should support the development of interactive explanations since communication technology, e.g., pictures, can, in principle, remain the same.
\citet{putnam2019exploring} presents one possibility for designing interactive explanations for intelligent tutoring systems,
where users can receive more detailed explanations by asking the system why this happened or how it happened.

For mobile health apps, we investigate three different variations of interactive explanations (i.e., live explanation, feature explanation, and ask-the-app explanation). 

\section{Persona Concept}
\citet{holzinger2022personas} emphasize the importance of asking stakeholders about their attitudes towards AI. This information is essential, as it influences whether and how users will utilize an AI system later.
Personas represent fictional stakeholders \cite{sim2014empowering}; they help developers understand their target users, empathize with them, and make better decisions concerning the usage of a system. \citet{cooper1999inmates} was one of the first to present this concept as a part of requirements engineering to focus on who uses the system.
The literature proposes different templates for the guided construction \cite{ferreira2018technique}. Usually, these are textual descriptions that include information about the person’s background, behaviors, and personal traits \cite{anvari2015effectiveness}. \citet{ferreira2018technique} identified two limitations concerning such persona templates. First, techniques neglect to elicit requirements to focus more on the empathy aspects. Second, specific survey methods do not relate to the application domain and provide general, less
context-specific characteristics for behavior identification.
To combine both, \citet{ferreira2018technique} propose the ‘PATHY 2.0’ technique (Personas empATHY).
Its purpose is to enable the identification of potential application requirements, which are derived from user needs. Among other things, it aims at creating profiles that represent system stereotypes.
The technique separates into six fields, each providing guiding questions to help describe the respective section: Who, context, technology experience, problems, needs, and existing solutions. 

\citet[p.7]{schneider2019personalized} describe that when XAI is being personalised, four relevant categories emerge:

\begin{itemize}
	\item \textbf{Prior knowledge} What does a user know?
	\item \textbf{Decision information} What information does the user want for the decision?
	\item \textbf{Preferences} What does the user like/prefer?
	\item \textbf{Purpose} What should the explanation be used for?
\end{itemize}

\begin{figure}[ht]
  \centering
	\begin{minipage}{\linewidth}
		\centering
		\includegraphics[width=0.65\linewidth]{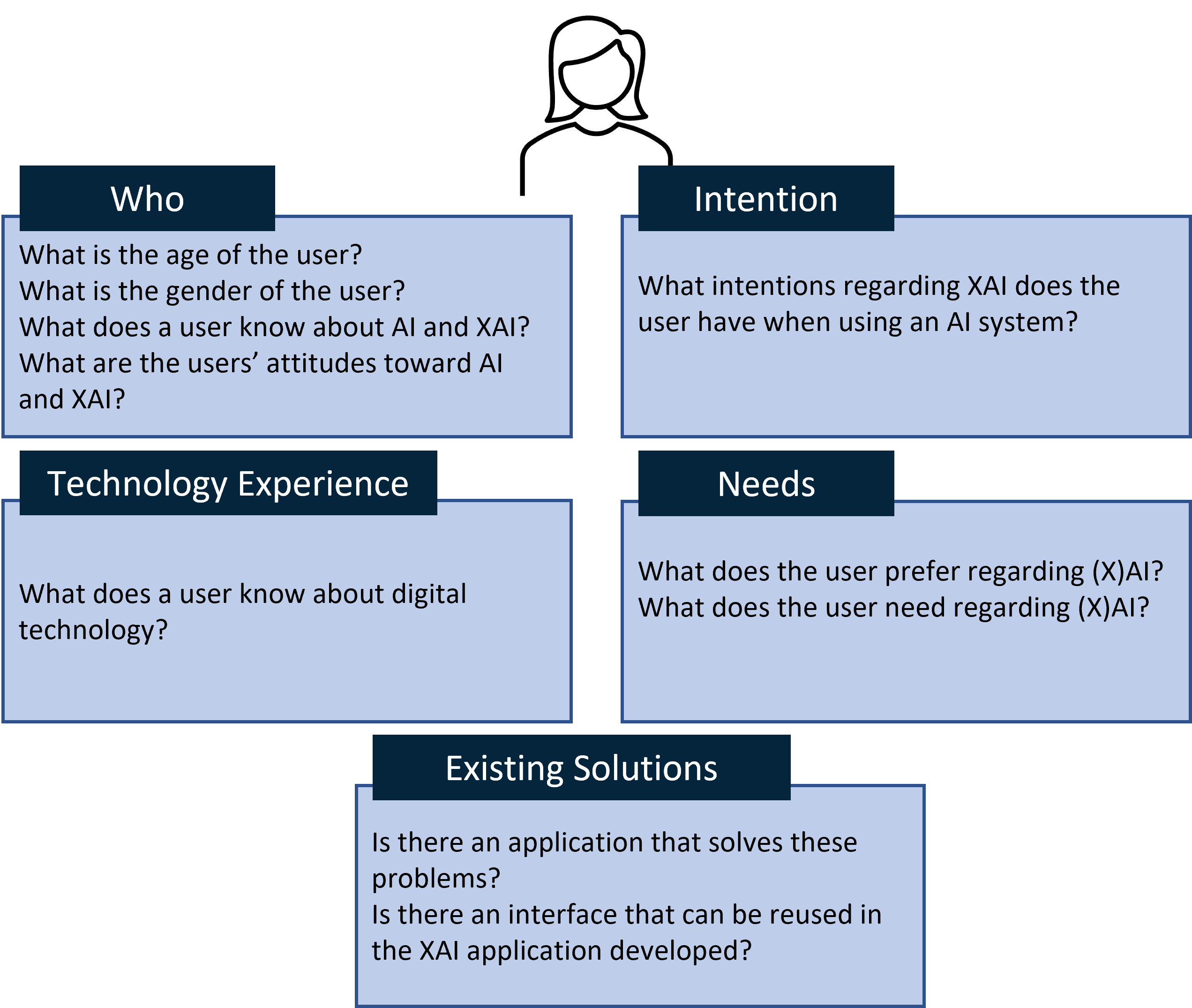}
	\end{minipage}
	\begin{minipage}{\linewidth}
		\centering
		\includegraphics[width=0.65\linewidth]{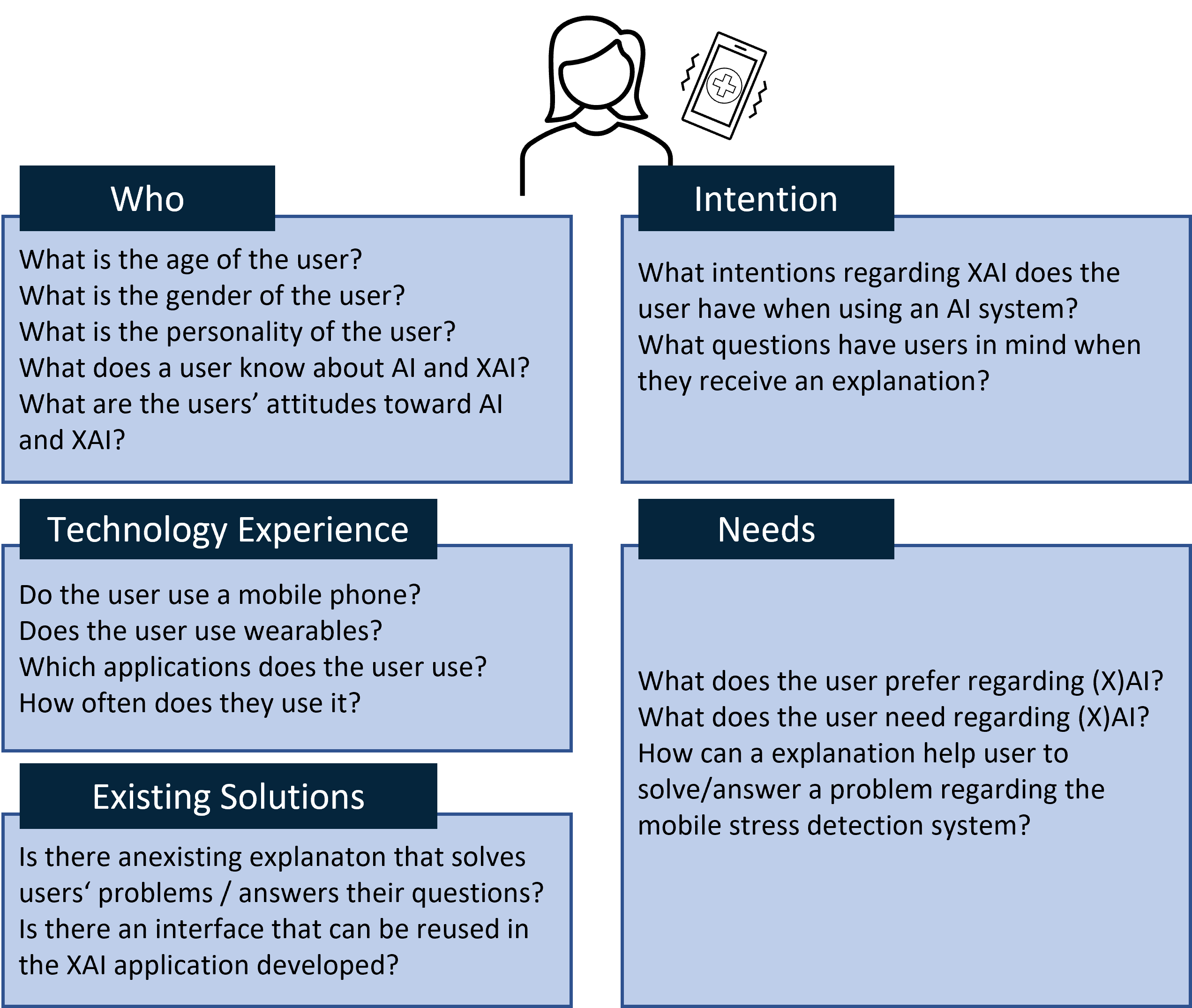}
	\end{minipage}
	\caption{\textit{Top}: Adapted persona template for XAI research, based on the work of \citet{ferreira2018technique,schneider2019personalized}. \textit{Bottom}: Persona template for the mobile stress recognition use-case in this paper}
	\label{fig:PersonaTemplate}
\end{figure}

To better reflect personalisation in the PATHY 2.0 approach, we included the recommendations of \citet{schneider2019personalized} in our human-centered XAI persona template. Further, we replaced `problems' with `intentions' as it appropriately fits personalised explanations. Lastly, we removed the `context field' as it primarily focuses on daily life and routines. This is out of the scope of this paper as the focal point is on the human-grounded evaluation and does not fully immerse into an application-grounded evaluation for which such context would be necessary \cite{ferreira2017identifying}.
These adoptions result in a modified version of the PATHY 2.0 technique for human-centered XAI, seen in \Fref{fig:PersonaTemplate}. 

The following displays the used template and a corresponding summary of each section:

\begin{itemize}
    \item \textbf{Who} Description of who will use the explanation or application. This field introduces users, their traits, frustration, and concerns.
    \item \textbf{Technology experience} Describes user experiences with other applications
    from this domain or other explanations. Additionally, it names application or explanation
    preferences.
    \item \textbf{Intentions} Reasons why users use the application or explanation. This field’s goal is to illuminate user issues and their core problem.
    \item \textbf{Needs} It states the needs required to fulfill the intentions and problems stated in the last field (Intentions).
    \item \textbf{Existing solutions} A listing of existing solutions trying to solve a similar problem or are similar in their idea and approach to the designed system.
\end{itemize}


\section{Research Questions}
For investigating our human-centered XAI approach in combination with creating empirically driven personas, we chose stress detection monitoring with the help of DNN \cite{li2020stress} as a practice-relevant use case for the presented explanations.
Since digital health is one of the main areas for XAI appliances \cite{adadi2018peeking}, it challenges XAI with a broad user group, ranging from AI experts to clinicians and patients. \citet{tjoa2020medicalXAI} emphasized the need for improved explainability for these heterogeneous user groups. Limiting oneself to one explanation will restrict the use of such a diversified user base and, ultimately, the help provided through monitoring applications. This paper's scope includes the user study's evaluation and the resulting persona evaluation, which are used to refine initial approaches and obtain a first listing of the requirements. For this, three visual-interactive explanations were created (see \Fref{fig:MTurk-XAITypesOverview}), and their usefulness for users of a stress recognition AI system was investigated.
The central research questions (RQ) this paper would like to provide answers to are:

\begin{itemize}
	\item \textbf{RQ1: App \& Explanation Preferences}
	\begin{itemize}
		\item \textbf{RQ1a}: Which representation of an AI classification system (i.e., data-based or photo-based) do end-users prefer?
		\item \textbf{RQ1b}: Which content and type of explanation is preferred by end-users?
	\end{itemize}
	\item \textbf{RQ2: Impact of Users' Attributes} Are end-users attitudes towards explanations related to personality, technical affinity, or demographic attributes (e.g., gender, age)?
\end{itemize}

\section{Online Survey}
We conducted an Amazon Turk (MTurk) online survey to investigate the research questions. Here, we addressed English and German-speaking people. On average, the survey took 30 minutes, and participants were compensated \$4.
The data protection officer of \textit{blinded for review} approved the online survey. At the beginning of the survey, participants were informed about the goals and duration of the study, and their GDPR rights.

\subsection{Survey Design}
The online survey comprised three phases:
\begin{itemize}
	\item \textbf{User information \& Preferences} At the beginning of the survey, we collected demographic information, personality traits, and users' attitudes regarding technical affinity and their attitudes and usage of health applications. In addition, we asked about participants' attitudes towards (X)AI only at the end of the survey to not bias users.
	\item \textbf{Mobile Health Application} Next, we presented the users with an example of a photo-based app and a data-based app and asked them about their preferences for one of the two apps. We then asked how much they would like an explanation of the app shown and what questions they would ask of the app.
	\item \textbf{Explainability} Based on their preference for one of the two mobile health apps, participants saw the input data of a potential person, which differed regarding the type of the app\footnote{For the data-based app, heart rate, blood oxygen, EDA, calendar entries, and sleeping time of the imaginary person has been displayed. For the photo-based app, a picture of a person was shown}. Participants had the task to classify this person as stressed / not stressed and how they would explain their decision (i.e., ``Assume that you are supposed to explain this decision to someone else. You may use all information the app provides.''). 
	\item \textbf{Personalised Explanations} In the last section of the survey, participants rated three different types of explanations (i.e., live explanation, feature-based explanation, and ask-the-app explanation)  and provided information about the time they would spend understanding the explanation. 
\end{itemize}

\begin{figure}[ht]
    \centering
	\includegraphics[width=0.97\textwidth]{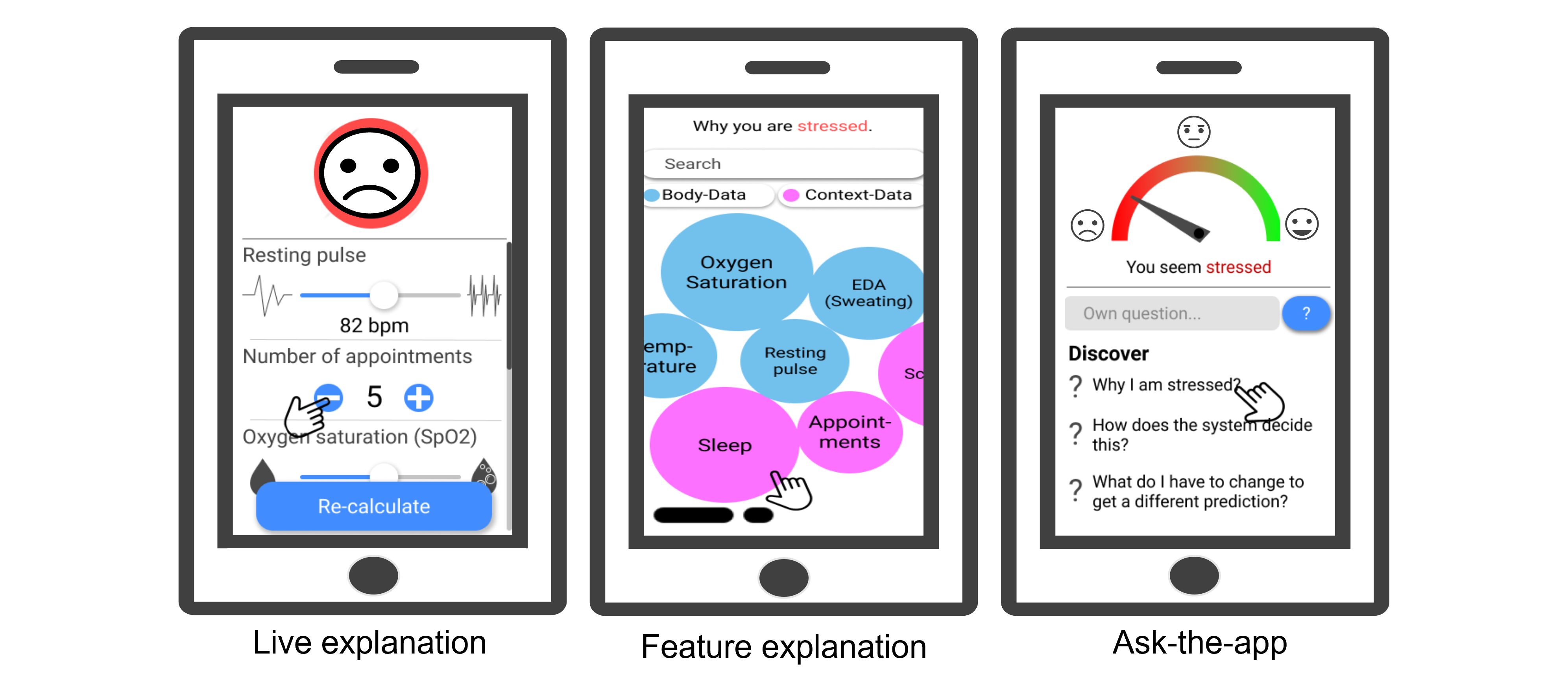}
	\caption{Three types of interactive explanations when classifying stress/no stress were investigated in the survey. Images illustrate the explanations for the data-based app: \emph{Live explanation} that allows users to change parameters, \emph{Feature explanation} that displays the features used for the decision, where the size of the circle report the importance of the feature for the classification, and \emph{Ask-the-app} explanations where users can ask the app-specific questions regarding the decision. The hand icon symbolizes a user action}
	\label{fig:MTurk-XAITypesOverview}
\end{figure}

We used three different kinds of explanation designs to investigate personalized explanations: live explanation, feature-based explanation, and ask-the-app explanations (see \Fref{fig:MTurk-XAITypesOverview}). All three explanation types have in common that the user can actively influence the explanation. This makes the explanation system interactive \cite{weld2019challenge}. However, since we were interested in user preferences and less in concrete AI models, the explanations shown are not based on real AI models. 

\begin{itemize}
    \item \textbf{Live explanation} The live explanation implements an exploratory explanation paradigm \cite{shneiderman2020trustworthyHCAI}. The idea is that the user finds out how input variables (e.g., resting pulse) affect output (e.g., stress classification) by trial and error. 
    \item \textbf{Feature explanation} A feature tag approach was chosen for this type of explanation\footnote{Here, data can be displayed in different sizes and colours indicating, for example, different degrees of importance \cite{halvey2007assessment}.}. In our explanations, different features of a (fictional) ML model are used. The size of the displayed feature represents the influence of the feature on the prediction. More detailed information about a feature is displayed to the user by clicking on it.
    \item \textbf{Ask-the-app explanation} This explanation is a dialogue-inspired explanation. This type of explanation allows the user to ask questions about the prediction of the AI system or to choose from several predefined questions, the one whose answer interests him.
\end{itemize}

\subsection{Methodology}
The survey included the following questions:

\paragraph{Personality} To assess the personality of the participants, we used the Ten-Item Personality Inventory (TIPI) questionnaire developed by \citet{gosling2003very}. TIPI investigates the Big-Five personality construct (i.e., extraversion, agreeableness, conscientiousness, emotional stability, openness to experience). For this, ten items (e.g., ``I see myself as sympathetic, warm`` for agreeableness) on a 7-point Likert scale (1 = Disagree strongly to 7 = Agree Strongly) were asked. 

\paragraph{Technical Affinity} To measure the technical affinity of the participants, we used the Affinity for Technology Interaction Short scale (ATI-S) \cite{wessel2019ati} which includes five items less than the original ATI scale \cite{franke2019personal}. Each of the four ATI-S items (e.g., ``I like testing the functions of new technical systems'') was rated on a 6-point Likert scale (1 = completely disagree to 6 = completely agree).

\paragraph{Health App} We investigated participants' usage of mobile devices for healthcare applications. Afterward, participants were asked which of the two apps presented (i.e., data-based app or photo-based app) they would use and then had to justify their preferred app via free-form feedback. 

\paragraph{Explanations - Type} Participants had to rate three different types of explanations (see \Fref{fig:MTurk-XAITypesOverview}). For this, three visualisations were evaluated: \emph{live explanation}, \emph{feature-cloud explanation}, and \emph{ask-the-app explanation}. Here, we used five items of the Explanation Satisfaction Scale (ESS) proposed by \citet{hoffman2018MetricsXAI}: understanding, satisfaction, sufficient detail, useful for users' goals, and precision (rating from 1 = I disagree strongly to 5 = I agree strongly). In addition, we asked participants about their willingness to try the explanation and their intention to personalize the explanation (e.g., ``I would like to determine for myself what factors are considered for the ask-the-app explanation.'') on a 5-point Likert scale (1 = I disagree strongly to 5 = I agree strongly). After that, participants could state for each explanation type what they liked or disliked, using free-form feedback.

\paragraph{Explanations - Content} Regarding the content of explanations, users were asked to rate which of the four presented questions they would ask themselves (i.e., ``How likely would you be to ask yourself any of the following questions while using the app?'') on a 5-point Likert scale (1 = extremely likely to 5 = extremely unlikely). The questions presented are inspired by \citet[p. 4]{hoffman2018MetricsXAI} and \citet[p. 2120]{lim2009and} and are intended to map user needs to the app and the explanations required:

\begin{itemize}
	\item Why do I get this prediction?
	\item How does the system come up with this prediction?
	\item Why did I not get another prediction?
	\item What do I have to change to make the system change its prediction?
\end{itemize} 

In addition, we investigated participants' preferred content of explanation (i.e., level of detail, comparison with the average of users) by contrasting two explanations (e.g., explanation with few details, explanation with many details) and having participants rate which of the explanations they would prefer. 

\paragraph{Explanation - Time}
To investigate the amount of time participants are willing to spend for explanations, we asked two questions: (1) How much time would you invest in understanding the app's explanation?'' (less than 1 minute; 1-2 minutes, 2-5 minutes, more than 5 minutes) and (2) ``If you had the opportunity to ask questions to the app or interact with the app, would you be willing to put more time into understanding an explanation?'' (5-point Likert scale; 1 = totally disagree, 5 = totally agree)

\section{Participants}
A total of 92 participants between 24 and 70 years (\textit{M}~=~ 42.10, \textit{SD}~=~10.4) finished the online survey. Forty-two of the participants were female, 49 male, and one diverse. All of the participants stated that they are currently located in the United States. 35.8\% of the participants indicated that they had a secondary degree or apprenticeship degree, 64.1\% stated that they have a university degree. 

68.6\% of the participants use a mobile phone in their daily life, and 28.3\% use a mobile phone in combination with a wearable. 50\% of the participants reported using health-related apps at least once a week. Especially the usage of fitness apps (\textit{n} = 44), followed by well-being apps (\textit{n} = 23), and nutrition apps (\textit{n} = 23) was reported.

\begin{figure}[ht]
    \centering
	\includegraphics[width=0.5\linewidth]{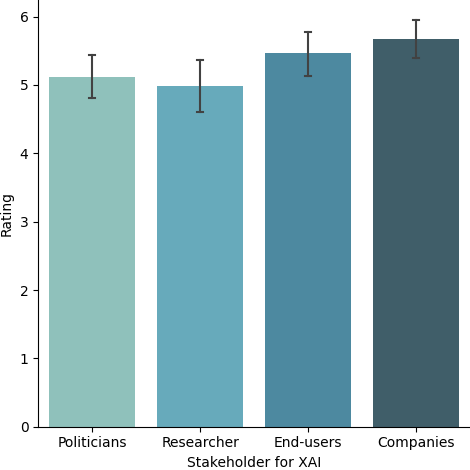}
	\caption{Rating of the users regarding the importance of XAI. Participants stated that XAI is important for different stakeholders, especially for end-users and companies}
	\label{fig:MTurk_XAIImportance}
\end{figure}

97.8\% of the participants stated that they had heard the term AI and 96.7\% of them stated that they agreed with the definition of AI given\footnote{The definition presented was: ``The term `artificial intelligence' is often used to describe machines (or computers) that mimic 'cognitive' functions that humans associate with the human mind, such as ``learning'' and ``problem-solving'' and is oriented on the definitions given by \citet{russell2016artificial}}. 95.7\% of the participants stated that they had a slightly or strong positive attitude towards AI (rated 4 or more on a 7-point Likert scale).
Regarding XAI, 95.7\% of the participants were not familiar with the term. Nevertheless, after giving a definition of XAI that focuses on the purpose of XAI\footnote{XAI was described to all participants as: ``With the help of explainable artificial intelligence, it should be possible to have a better understanding of artificial intelligence''}, 93.5\% had a slightly or strong positive attitude towards XAI (rated 4 or more on a 7-point Likert scale). Participants stated that XAI is important for different stakeholders, especially end-users and companies (see \Fref{fig:MTurk_XAIImportance}).

\paragraph{Technical Affinity} Participants had a technical affinity of \textit{M}~=~4.03, \textit{SD}~=~1.14 with a Cronbach's $\alpha$ of .88\footnote{\citet{wessel2019ati} suggest to report Cronbach's $\alpha$ as an indicator of reliability. Here, values between 0.7 to 0.8 are indicating a satisfactory reliability \cite{bland1997statistics}}. The results of a one-sample t-test shows that participants had significant high values in technology affinity, \textit{t}(91)~=~4.47, \textit{p}~$<$~.001, \textit{d}~=~.47 (medium effect)\footnote{the effect size \textit{d} is calculated according to \citet{cohen2013statistical}. Interpretation of the effect size is: \textit{d}~$<$~.5~:~small effect; \textit{d}~=~0.5-0.8~:~medium~effect; \textit{d}~$>$~0.8~:~large~effect} compared to the mean of 3.5 (6-point Likert scale).
The technical affinity did not significantly correlate\footnote{we calculated Spearmans' rang correlations} with age nor gender.

\paragraph{Personality} 
Regarding the TIPI's Big-5 items, the scales agreeableness and conscientiousness had a very low Cronbach's $\alpha$, indicating low reliability of these scales and were therefore excluded from further analyses (see \Fref{tab:MTurk-Health_Personality}.
\begin{table}
	\caption{Cronbach's $\alpha$ for the five personality items of the TIPI questionnaire \cite{gosling2003very}. Agreeableness and conscientiousness are below 0.7 and therefore indicate a not sufficient reliability}
	\label{tab:MTurk-Health_Personality} 
	\centering
	\begin{tabular}{lc}
		\hline\noalign{\smallskip}
		\textbf{TIPI item} & Cronbach's $\alpha$ \\
		\noalign{\smallskip}\hline\noalign{\smallskip}
		Extraversion & .81  \\
		Agreeableness & .42  \\
		Conscientiousness & .67  \\
		Emotional Stability & .84\\
		Openness to Experiences & .71 \\
		\noalign{\smallskip}\hline
	\end{tabular}
\end{table}

\section{Results}

\subsection{RQ1: Preferred App \& Explanations}
\subsubsection{RQ1a: Preferred App} 
With 88\%, most of the participants preferred the data-based app. Reasons for this could be investigated in the free-form answers. We found three general reasons why people prefer one of the two apps:

\begin{itemize}
	\item \textbf{Dependability}: Thirty-eight participants (3 of them preferred the photo-based app) stated several things regarding dependability. Here it was frequently mentioned that the data-based app is more reliable since it uses much more data for evaluation than the photo-based app (e.g., ``It would be able to collect data about me that would not be visible in a photo.''). In general, participants were very critical of the reliability and general functioning of the photo-based app (e.g., ``The other one [photo-based app] seems like pseudoscience.''), whereas the data-based app was perceived as more objective (e.g., ``Further, I think the data-based application is more likely to be objective, whereas deducting characteristics from a photo seems more subjective and therefore less valuable to me.'').
	
	\item \textbf{Privacy}: Twenty-six participants (1 of them preferred the photo-based app) raised privacy concerns. In particular, people did not feel comfortable providing a photo of themselves and therefore tended to use the data-based app (e.g., ``I don't like the idea of taking my photograph in an app, because I may not have control over how that photograph may be used.''). Regarding the data-based app, participants tend to assume that the shared information was not so personal compared to the photo-based app (e.g., ``I suspect that more personal information about myself could be obtained via the photo than via the data used in the data-based application.'').
	
	\item \textbf{Usability}: Twenty-two participants (5 of them preferred the photo-based app) stated reasons related to the app's usability. While the photo-based app was described as easier and quicker to use (e.g., ``It is quicker to understand and look over), the data-based app was described as more understandable (e.g., ``I feel like it gives me a better understanding of my health than my face'').
\end{itemize}

\subsubsection{RQ1b: Preferred Explanations}

\paragraph{Preferred Content of Explanation} Based on the triggers \emph{why, how, why not, and what} proposed by \citet{hoffman2018MetricsXAI}, we investigated which information participants want to know from a stress-classification app. Results of a one-sample t-test show that all questions except the Why not?-question were significantly positively rated and can therefore be assumed to be relevant for users in mobile health scenarios (see \Fref{tab:MTurk-ratingXAIQuestions}). In addition, users tend to prefer more detailed explanations (72.8\%) as well as explanations that include a comparison to the average (68.5\%). 

\begin{table}
	\caption{Rating of potential questions users would ask themselves when seeing a stress classification of a mobile health app. A one-sample t-test revealed that all question types except Why not?-questions were perceived as significantly important by participants}
	\label{tab:MTurk-ratingXAIQuestions} 
	\centering
	\begin{tabular}{lrcc}
		\hline\noalign{\smallskip}
		\textbf{Question type} & \textit{t}(91) & \textit{p} & \textit{d} \\
		\noalign{\smallskip}\hline\noalign{\smallskip}
		Why? & 9.26 & $<.001$** & 0.97 \\
		Why not? & 0.19 & .849 & 0.02 \\
		How? & 9.16 & $<.001$** & 0.95 \\
		What? & 3.75 & $<.001$** & 0.39\\
		\noalign{\smallskip}\hline
		**\textit{p}$<$.001
	\end{tabular}
\end{table}

\paragraph{Preferred Type of Explanation}
Overall, 56.8\% of the participants prefer to use the ask-the-app explanation in their daily life, followed by the feature explanation (23.5\%) and the live explanation (19.8\%).

\begin{figure}[ht]
    \centering
	\includegraphics[width=0.5\linewidth]{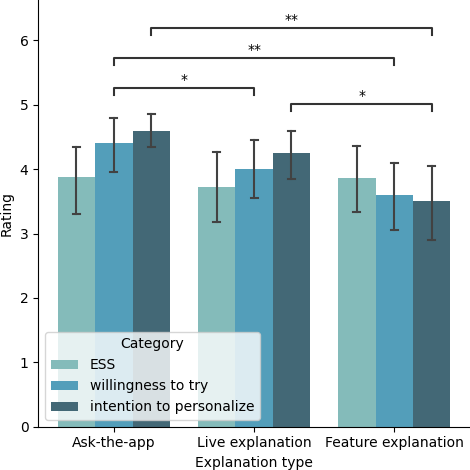}
	\caption{Participants rating (ranging from 1 to 7) of the three different interactive explanation types (i.e., ask-the-app, live explanation, and feature explanation) shows that ask-the-app invites users to try it out. Furthermore, ask-the-app and live explanations seem to invite users to personalize them (i.e., selecting features that should be relevant for the classification). Error bars represent the 95\% CI. *~\textit{p}~$<$~.05, **~\textit{p}~$<$~.001}
	\label{fig:MTurk-XAITypePreferred}
\end{figure}

Comparing the three \emph{types} of explanation (i.e., live explanation, feature-cloud explanation, and ask-the app explanation), we found that the explanation satisfaction did not differ significantly between the types (see \Fref{fig:MTurk-XAITypePreferred}). 

Regarding the willingness to try the explanation type, the ask-the-app explanation was rated significant higher than the live explanation, \textit{t}(91)~=~-2.42, \textit{p}~$=$~.018, \textit{d}~=~.25 (small effect) and than the feature explanation, \textit{t}(91)~=~-3.53, \textit{p}~$<$~.001, \textit{d}~=~.37 (medium effect). 
While the explanation satisfaction seems similar to each explanation type, the ask-the-app invites users to try it out - maybe because the app forces an interaction through asking questions (see \Fref{fig:MTurk-XAITypePreferred}). 

Regarding the intention to personalise the explanation, the live explanation was rated significant higher than the feature explanation, \textit{t}(91)~=~2.97, \textit{p}~$=$~.004, \textit{d}~=~.31 (medium effect). The ask-the-app explanation was rated higher than the feature explanation, \textit{t}(91)~=~4.29, \textit{p}~$<$~.001, \textit{d}~=~.45 (medium effect). It seems that the feature explanation is not so interesting for users to personalize (see \Fref{fig:MTurk-XAITypePreferred}).

\paragraph{Time Spend for Explanations} Regarding the interaction time, participants' answers were evenly distributed between 1 to 2 minutes up to more than 5 minutes (see \Fref{tab:MTurk-XAITimeSpend}). The results indicate that users are willing to spend some time understanding an explanation. 

\begin{table}
	\caption{End-users answer of the question ``How much time would you spend to understand an explanation in the mobile health app?''}
	\label{tab:MTurk-XAITimeSpend} 
	\centering
	\begin{tabular}{lr}
		\hline\noalign{\smallskip}
		\textbf{Time} & \textbf{Percent}  \\
		\noalign{\smallskip}\hline\noalign{\smallskip}
		less than 1 minute & 6.5\% \\
		1-2 minutes& 30.4\%  \\
		2-5 minutes & 32.6\% \\
		more than 5 minutes & 30.4\%\\
		\noalign{\smallskip}\hline
	\end{tabular}
\end{table}

When asking whether participants are willing to spend more time with an explanation when asking questions or interactively interact with the app, we found a statistically significant difference to the mean of 3 (5-point Likert scale) conducting a one-sample t-test, \textit{t}(91)~=~11.5, \textit{p}~$<$~.001, \textit{d}~=~1.19 (large effect), indicating that users tend to spend more time with an app that provides an interactive interface for explanations. 

\subsection{RQ2: Impact of Users' Attributes}
We will report correlations regarding user attributes and the explanations presented in the following. However, the reported values only reflect correlations, not causal relationships.

\paragraph{Demographic Information}
Regarding the time participants would spend to understand an explanation, we found a significant positive correlation with the participants' age (\textit{$r_{sp}~=~.24, \textit{p}~=~.019$}), indicating that the older the users, the more willing they are to spend time to understand an AI's explanation. No significant correlations were found for gender and educational background.
Regarding time participants would spend on an explanation when they have the option to ask questions during the explanation, we found a significant positive correlation with the participants' attitude towards AI (\textit{$r_{sp}~=~.30, \textit{p}~=~.004$}) and XAI (\textit{$r_{sp}~=~.28, \textit{p}~=~.008$}). A more positive attitude towards AI as well as XAI leads to higher time investments to understand an explanation. 

Regarding attitude towards AI and XAI, we only found a significant positive correlation between gender and the attitude towards AI (\textit{$r_{sp}~=~.23, \textit{p} = .026$}), indicating that men have a more positive attitude towards AI. For attitude towards XAI, no significant impact on age and gender was found. 
Regarding educational background, we found no relationship to participants' attitudes toward (X)AI nor their willingness to spend more time with an explanation or to ask questions to an explanatory system. 

These results indicate that the attitude towards (X)AI seems to be an essential driver for users to take the time to interact with an explanatory system. Gender and age seem to have an impact on the attitude towards AI, but not on XAI. Also, the educational background does not indicate the attitude towards (X)AI in general nor the interaction with an explanatory system. 

\paragraph{Personality}
To evaluate the impact of personality on the different explanation ratings, due to Cronbach's $\alpha$, only the traits \emph{Extraversion}, \emph{Emotional Stability}, and \emph{Openess to Experiences} are used. 

Regarding time participants would spend on an explanation when they have the option to ask questions during the explanation, we found a significant positive correlation with the personality trait Extraversion (\textit{$r_{sp}~=~.22, \textit{p}~=~.036$}) and Openness to Experiences (\textit{$r_{sp}~=~.23, \textit{p}~=~.030$}). This indicates that users with higher values in these two personality traits are more willing to spend time with an interactive explanation. 

Regarding the time participants would spend understanding an explanation, we found no significant relationship with any personality traits. 

Regarding the content of the explanation (i.e., comparison to average user and degree of detail of the explanation), we found no significant correlation with any of the personality traits. 

\paragraph{Technology Affinity}
Regarding the time participants would spend to understand an explanation, we found a signification positive correlation with the technical affinity of the participants (\textit{$r_{sp}$}$~=~.27$, \textit{p} = $.010$), indicating that users with higher values in technical affinity tend to invest more time in understanding an explanation. 

Regarding the content of the explanation (i.e., comparison to average user and degree of detail of the explanation), we found no significant correlation with technology affinity. 

\section{Derived Personas}
\begin{figure}[ht]
    \centering
	\includegraphics[width=0.92\linewidth]{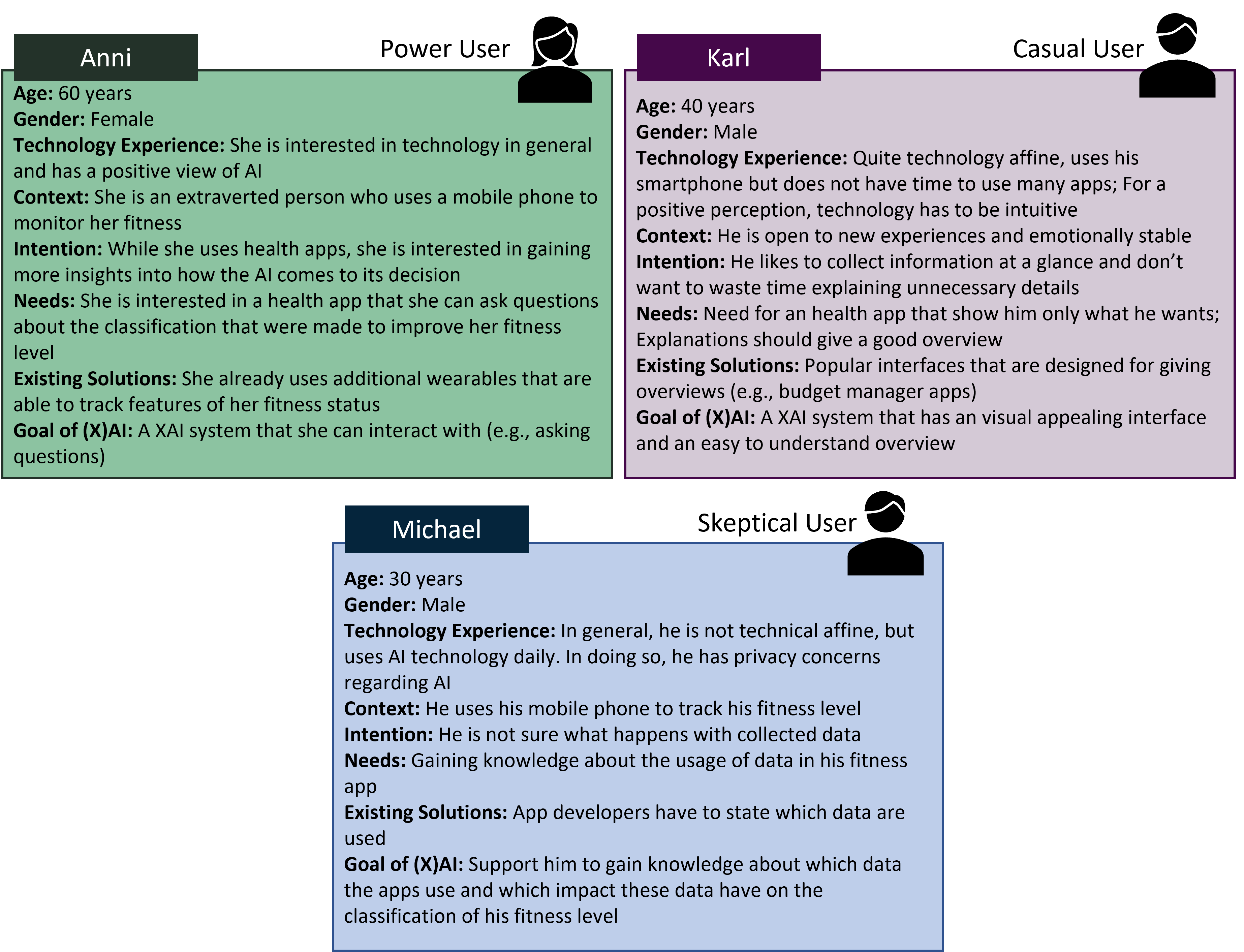}
	\caption{Three personas were derived based on the empirical data: \emph{Anni}, a power user, \emph{Karl}, a casual user, and \emph{Michael}, a skeptical user}
	\label{fig:DerivedPersonas}
\end{figure}

Building on the findings collected during the survey and the requirements derived, we designed user personas in the next step. The different argumentative identified during the application selection has indicated first divergences in the motivation and attitude of the users. We found three different views of users: Dependable and accuracy focused (Persona 1 - Anni), perception and usability focused (Persona 2 - Karl), and privacy and commitment focalized (Persona 3 - Michael) (see \Fref{fig:DerivedPersonas}).

\subsection{Clustering of Results}
\begin{figure}[ht]
    \centering
	\includegraphics[width=0.8\linewidth]{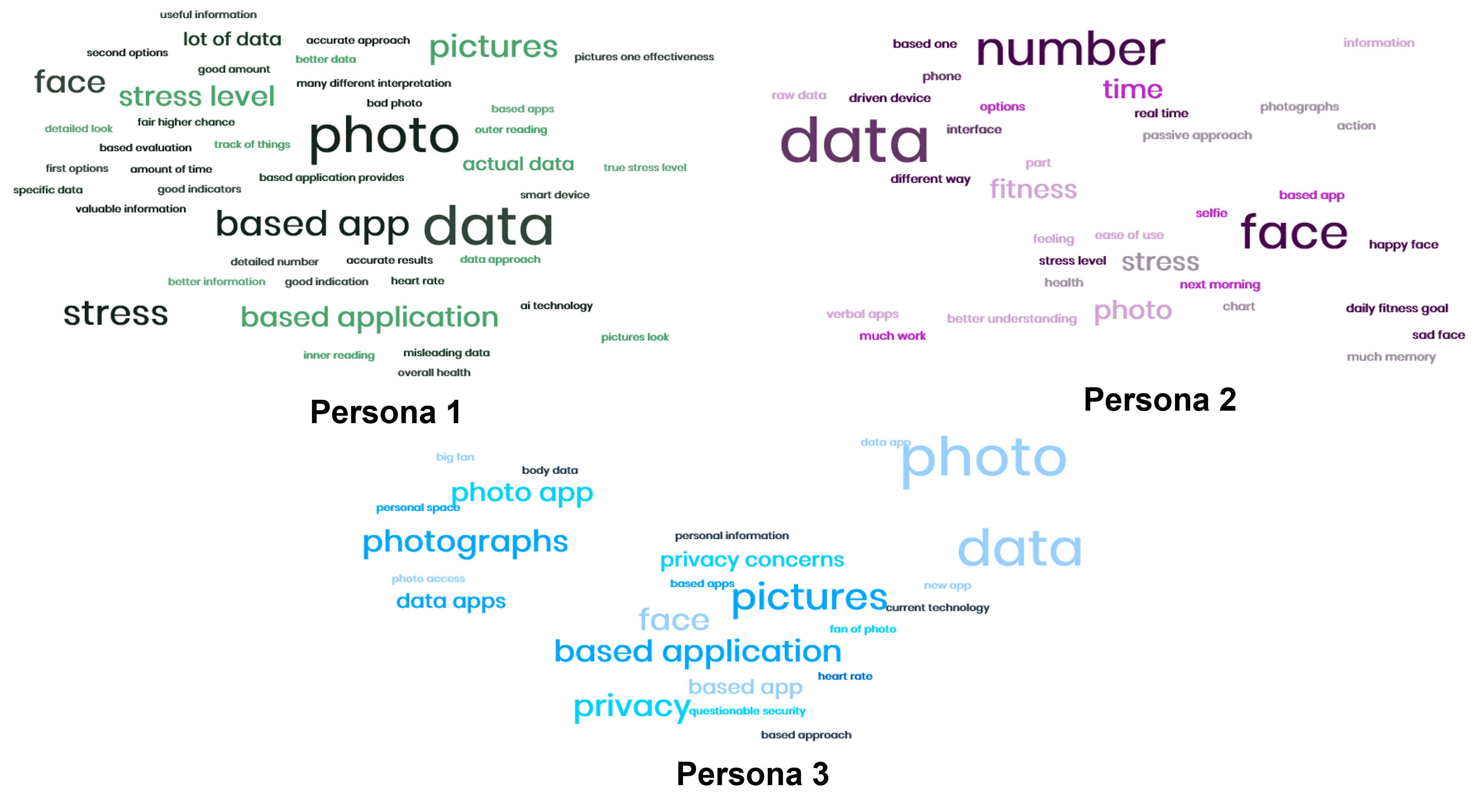}
	\caption{We generated word clouds for the three different clusters of prototypical users based on the free-form feedback to the question ``Why do you prefer this app over the other?''. The overall sentiment for persona 1 and 2 was positive, while the sentiment for persona 3 was negative}
	\label{fig:WordCloudPersonas}
\end{figure}

The different argumentative identified during the online survey have indicated first divergences in the motivation and attitude of the users. The results regarding \textit{RQ1a: Preferred App} introduced three views: Dependable and accuracy-focused users (persona 1), perception and usability-focused users (persona 2), and privacy and commitment focalized users (persona 3).
With these three clusters as a basis for three personas, additional answers were investigated for possible patterns and affirmations of the personas. Due to the low number of respondents who chose the photo option, the results are based on the proportion of the data-based app.
The positive and negative impressions for the different explanation types were now separated according to the three initial clusters (see \Fref{fig:WordCloudPersonas}). Analysing users' free form feedback content, we found an overall positive sentiment for persona 1 with 99.4\% and persona 2 with 98.5\%. However, for persona 3, we found 96.1\% an overall negative sentiment, reflecting the aversion against the photo-based app.

Next, we summed up general requirements found during the analysis of the results into \Fref{tab:RequirementsXAI}. The requirements were elicited
from the participants’ preferences, intentions, and feedback for the three interactive explanations shown.

\begin{table}
	\caption{XAI requirements and their description. Requirements were elicited from the user survey}
	\label{tab:RequirementsXAI} 
	\centering
	\begin{tabular}{ll}
		\hline\noalign{\smallskip}
		\textbf{Requirement} & \textbf{Description}  \\
		\noalign{\smallskip}\hline\noalign{\smallskip}
		Simplicity & The interface should be easy and fast to understand \\
		Forecast& To see the impact of a parameter if its value would be changed  \\
		Quality & Explanations should provide sufficient details \\
		Queryable & It should be possible to ask follow-up or personal questions\\
		&towards the AI\\
		Personalized & The contents and the explanation should be personalised \\
		&towards the recipient. Personalised explanations can \\
		&consume more time\\
		Context provision & Users should be able to provide additional context for the \\
		&measured features\\
		Feature explanations & Features used in the prediction should be explained\\
		\noalign{\smallskip}\hline
	\end{tabular}
\end{table}

\begin{figure}[ht]
    \centering
	\includegraphics[width=\linewidth]{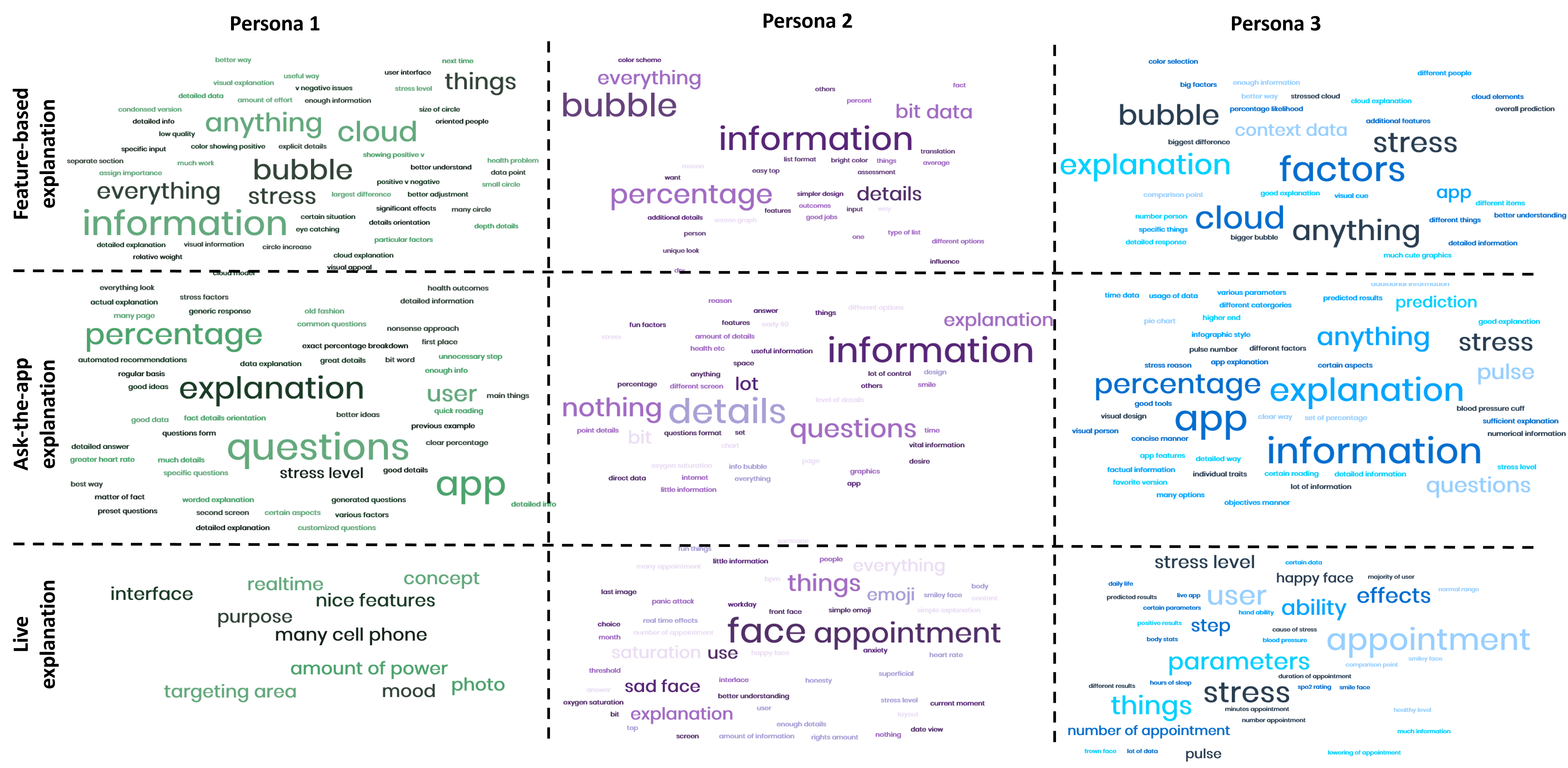}
	\caption{Word clouds regarding the three different types of interactive explanations (i.e., feature-based explanation, ask-the-app explanation, and live explanation) for the three personas. Bigger words indicate often mentioned terms}
	\label{fig:WordCloudThreeExplanationsPersonas}
\end{figure}

After that, we investigated how users of the three clusters responded to the three interactive explanation types (i.e., ask-the-app, live, and feature explanation). A summary of the free-form feedback for the three explanation types is displayed in \Fref{fig:WordCloudThreeExplanationsPersonas}. Comparing the different explanation types at a glance, all users mention similar terms. 

Our results for \textit{RQ1b: Preferred Explanations} paint a more precise picture. Based on the closed-ended questions, the \textit{ask-the-app explanation} was generally perceived as the most satisfying. However, the personas express different tendencies of satisfaction. Persona 1 and
Persona 3 mentioned the improved depth and amount of details the most. Thus, both positively perceived the feature of asking questions. Conversely, Persona 2 focused on engaging design and understandability while criticizing the ask feature, as it is indirect and time-consuming.
Persona 3 also criticizes the high head complexity that the app entails.
For the \textit{feature-based explanation}, Persona 2 appreciated the ability to see the features’ impacts directly the most. Persona 1 is fond of the overview it offered, but at the same time, with 75\%, it was also the one who criticized the lack of comprehensibility and usefulness the most. Persona 3 paints a balanced picture with the highest liking for the overall design, while superficiality and lack of depth are the most criticized.

From this more qualitative analysis, we further explored the persona types quantitatively.
27 users reflecting Persona 2 report a significantly higher emotional stability (\textit{M} = 5.80, \textit{SD} = 1.44) than the
34 of Persona 1 (\textit{M} = 5.25, \textit{SD} = 1.36) and the 26 of Persona 3 (\textit{M} = 4.54, \textit{SD} = 2.09), \textit{p} = .045. Furthermore, 34 participants identified as Persona 1 rated why, how, and why not question significantly higher in the probability of self-questioning than others assigned to different personas.

The question of what would have to change did not indicate any significant differences dependent on different personas as Persona 1 (\textit{M} = 3.79, \textit{SD} = 1.20), Persona 2 (\textit{M} = 3.19, \textit{SD} = 1,24), and Persona 3 (\textit{M} = 3.77, \textit{SD} = 1.31) demonstrated similar interests.

\subsection{Persona Description}

Persona 1 (Anni) and Persona 3 (Michael) placed the most value on the changeability and exploratory character of the interactive explanations. In contrast, Persona 2 (Karl) valued the simplicity and understandability of the design more and the ability to provide a good overview. Persona 1 criticized the lack of details and inability to be more specific about the feature measured the most. Persona 2 added the criticism of not showing enough information simultaneously.

\paragraph{Persona 1: Power User}
Based on the qualitative and quantitative analysis of the data collected following statement was derived for the first persona: \textbf{Power user Anni, who enjoys details and technology, willing to put significant effort into applications found beneficial.} According to the Oxford Learners' Dictionary, the term power user refers to someone able to use more advanced features and engage in more complex topics than other users \cite{oxford2022}. 
Anni is intended to mirror users in our survey who actively use mobile phones to track their health, e.g., to improve their fitness. An interactive XAI design is the basis to satisfy Anni's request for queries to the app that encapsulates these traits. 
Anni is described as a persona with high precision standards and unsatisfied with uncertainties or vague statements. According to the data, her dominant personality traits are diligence, tolerance, and emotional stability. Further, the data shows Persona 1 as engaging in technology and an active user of mobile health apps. Therefore, practical explanations should extend her knowledge and allow her to deepen her understanding of the respective field. In contrast, unsuitable explanations lack these features.

\paragraph{Persona 2: Casual User}
The following persona statement can be derived from the previous results for the second persona: \textbf{Casual user appreciates easy and fast consumable information and does not want to spend too much time retracing AI.} Data on Persona 2, \emph{Karl}, displayed temperate interest in details while
understanding the core principles of the prediction, why, and how it was derived are essential. A deeper immersion is often not perceived as necessary by casual users. Hence, simplicity, understandability, as well as general appearance are valued. Due to the distribution of Persona 2 in the personality traits, his most pronounced traits are emotional stability, diligence, and openness to new experiences. He uses a smartphone and monitoring apps but is less intensive than Anni (Persona 1). Persona 2 is more usability-focused. Hence, an intuitive design is vital for Karl. This type displays reduced interest in intentions besides why and how. Existing solutions that focus on providing overviews can serve as suitable orientations.

\paragraph{Persona 3: Skeptical User}
Our online survey showed that the application area of healthcare triggers security concerns among users due to the collection of sensitive health data. The persona \emph{Michael} is prototypical for this critical user group. This last type is influenced by hesitancy and general usage concerns. Therefore, the subsequent persona statement was constructed: \textbf{Skeptical user, who is reserved about sharing too much information with new applications, commitment increases with increased trust}. Michael is described as somewhat skeptical and reserved. Privacy and nontransparent applications agitate him. Thus, he uses trusted applications and limits monitoring apps on non-invasive systems. Since Persona 3 is initially very reserved, too forceful or manipulative explanations are rejected. The commitment could gradually increase with Michael gaining more trust in the application. Personalisation turns out to be more a means for incremental learning than for instant adaptation of explanations. Apps implementing such a learning curve can serve as valuable inspirations. This persona stood out for its concern and reluctance to divulge personal data. For example, many users chose the data-based application over the photo-based approach since a photo of one’s face was seen as too deep an invasion of privacy. The concerns affect the use of a mobile health app. XAI design could address these concerns and aim to increase transparency and promote user trust in the app. 

\section{Discussion}

\paragraph{Data-based XAI Applications are Preferred} The survey results show that participants preferred the data-based health app. Reasons for this were the \emph{dependability} and \emph{usability}. In addition, \emph{privacy} was mentioned by participants as an important reason for choosing the data-based app. The reason for this was mainly the argument that pictures are very personal information and that they do not want to share pictures that show their vulnerability (i.e., being stressed) with a system where they do not know if these pictures will be used only for this purpose. In these statements, it can be seen that XAI designers should be aware of users' feelings since they are powerful drivers of decisions. \citet{arrieta2020XAIconcepts} indicate that privacy awareness should be created with the help of XAI. Further research should investigate whether privacy concerns about a healthcare application persist even if they can be explained. 

\paragraph{Users Want Detailed Explanations} Similar to the results of pilot studies conducted in \citet{hald2021anerror} and \citet{weitz2021ourfault}, users stated in the survey that they want very extensive information within an explanation. However, this statement must be taken with a grain of salt because it makes a difference whether users are expressing their desires for explanations for a potential app or are being asked to evaluate the explanations of an app they are actually using. Whereas users in the pilot study of \citet{weitz2021ourfault} indicated that they wanted extensive and detailed information in an explanation, the main study showed that detailed explanations were quickly found to be annoying because (1) information was repeated and (2) the extent of the explanation was often disruptive to actual task completion because too much information impacted users. Similar were the results found in \citet{hald2021anerror}. While participants stated that they wished for explanations with a suggestion of solutions, in the main study, these explanations did not improve participants' impression of a VR robot.
For future studies, it would be helpful to apply our persona approach in such use cases to better match users' preferences when using an AI system.

\paragraph{Interaction and Personalisation are Important for Users} The survey results show that users prefer personalisation of explanations and the possibility of using an interactive explanatory system. For these aspects, users would also invest more time in dealing with the explanatory system. While explanation satisfaction is similarly high for all types of explanations, there are differences in the willingness to try out such a system as well as the personalisation by users. The explanation types ask-the-app and live explanation show the highest approval by users. These types are also the ones that allow users to interact with the system, such as asking questions or changing parameters that are relevant to stress. In addition, participants would ask themselves a lot of questions (i.e., Why?, Why not?, How?, What?) when getting a stress classification decision. 
These empirical results support the claim of \citet{shneiderman2020trustworthyHCAI}, who uses the example of an interface for mortgage loan explanations, which allows users to interact with the system and to try out and experience for themselves the results that arise from changes to parameters. The interactivity of an explanatory system is desired by users and must be considered in the XAI design of future studies.

\section{Conclusion}
AI in the medical sector offers the possibility of supporting clinicians and patients in the treatment process. One specific application is the use of mobile health applications. These are easy to use on one's smartphone or wearables and allow users to monitor and, if necessary, improve their state of health. However, a major criticism when using such applications, which usually work with ML, is the lack of explanation of such black-box systems. While several XAI methods have addressed this issue, there is still a lack of empirical research on the usefulness of such techniques for end-users. End-user refers to people with no expertise in ML and AI and little or no domain knowledge (e.g., medical training). Human-centered AI adapts AI systems to different stakeholders, such as end-users. This paper presented a human-centered persona concept that creates prototypical user groups for XAI based on empirical user data. This approach was applied in an online survey of end-users preferences for different explanation styles and content in a mobile health stress monitoring application. The survey results show that many end-users preferred a data-based app to a photo-based app.
Regarding personalisation, some users preferred more detailed explanations, while others preferred more intuitive, clearly designed explanations. Other users, on the other hand, were particularly concerned about protecting their data. The study's results served as the basis for three prototypical personas: a power user, a casual user, and a privacy-oriented user. Our approach and insights from a user-centered perspective offer support for developers of XAI systems to design them in a human-centered way; thus, our work brings an interactive, human-centered XAI closer to practical application.  

\bibliographystyle{unsrtnat}  
\bibliography{literature.bib}

\end{document}